\documentstyle[aps,preprint,epsfig]{revtex}

\tightenlines

\begin{document}
\title{Linear Response Calculations of Spin Fluctuations}
\author{S. Y. Savrasov}
\address{Max-Planck-Institut f\"ur Festk\"orperforschung, Heisenbergstr. 1, 
70569 Stuttgart, Germany.}
\date{May 1998}
\maketitle

\begin{abstract}
A variational formulation of the time--dependent linear response based on
the Sternheimer method is developed in order to make practical {\it ab initio%
} calculations of dynamical spin susceptibilities of solids. Using gradient
density functional and a muffin--tin--orbital representation, the efficiency
of the approach is demonstrated by applications to selected magnetic and
strongly paramagnetic metals. The results are found to be consistent with
experiment and are compared with previous theoretical calculations.
\end{abstract}


%

Full wave--vector and frequency dependent spin susceptibility $\chi $ is a
central quantity in understanding spin fluctuational spectra of solids. Its
knowledge accessible directly via neutron--scattering measurements is
important due to significant influence of spin fluctuations to many physical
properties and phenomena \cite{Moriya}, such, {\it e.g.}, as the electronic
specific heat, electrical and thermal resistivity, suppression of
superconductivity for singlet spin pairing, {\it etc }. In magnetically
ordered materials, transverse spin fluctuations are spin waves whose energies 
and lifetimes are seen in the the structure of transverse susceptibility.
High--temperature superconductivity, a highly exciting
phenomenon, whose origin is still not recognized, can be due to a spin
fluctuational mechanism \cite{Scalapino}.

Despite large past efforts put on the development of methods for {\it ab
initio }calculations of the dynamical spin susceptibility based either on
the random--phase--approximation (RPA)\ decoupling of the Bethe--Salpeter
equation \cite{Cooke1973}, or within density functional formalism \cite
{Callaway1975,DFT}, quantitative estimates of $\chi $ with realistic energy
bands, wave functions, and self--consistently screened electron--electron
matrix elements are scarce in the literature \cite
{Comm1,CookeNiFe,CallawayNi,WinterPdV}. This is not only due to the absence
of complete theory for the proper description of exchange--correlation
effects which is a true many--body problem, but also because standard
perturbative treatment of an electronic response has serious problems
connected with the summation over high--energy states and matrix inversion.

This paper proposes a method which avoids the latter two problems. The
method is a time--dependent generalization of an all--electron Sternheimer
approach \cite{LR} 
which has been proved to be very efficient in {\it ab
initio} calculations of phonon dispersions, electron--phonon interactions
and transport properties of transition--metal materials including high--T$%
_{c}$ superconductors \cite{EPI}. The method employs a muffin--tin--orbital
representation \cite{OKA} which allows to greatly facilitate the treatment
of localized states such, {\it e.g.}, as d-- and f--electrons of strongly
paramagnetic and magnetic materials whose studying is the main purpose of
this work.

Applications to transverse spin fluctuations in Fe and Ni as well as
calculations of paramagnetic response in Cr and Pd demonstrate an efficiency
of the approach and resolve some discrepancies found in previous
theoretical studies. In particular, experimental evidence of an optical
spin--wave branch for Ni \cite{ExpNi} and its absence for Fe \cite{ExpFe} is
correctly described by the present calculation which was not done in either
early semiempirical approaches \cite{CookeNiFe,CallawayNi}, or within a
recent frozen--magnon scheme \cite{Halilov}. For the
first time, the dynamical susceptibility is calculated {\it ab initio} for
paramagnetic Cr, a highly interesting material due to its incommensurate 
antiferromagnetism \cite{Cr}. The calculation predicts a wave vector 
of the spin density wave (SDW), and clarifies the role 
of Fermi--surface nesting. Strong long--wavelength spin
fluctuations of Pd are evident from the present and 
earlier \cite{WinterPdV} theoretical studies.

The description of the method starts by considering a small external
magnetic field 
\begin{equation}
\delta {\bf B}_{ext}({\bf r}t)=\delta {\bf b}e^{i({\bf q}+{\bf G}){\bf r}%
}e^{i\omega t}e^{-\eta |t|}+c.c  \label{f1}
\end{equation}
applied to a solid. Here $\delta {\bf b}=\sum_{\mu }\delta b^{\mu }{\bf e}%
_{\mu }$ shows a polarization of the field ($\mu $ runs over $x,y,z\,$\ or
over $-1,0,1$), wave vector ${\bf q}$ lies in the first Brillouin zone, $%
{\bf G}$ is a reciprocal lattice vector, and $\eta $ is an infinitesimal
positive quantity. If the unperturbed system is described by charge density $%
\rho ({\bf r})$ and, in general, by magnetization ${\bf m}({\bf r})$, the
main problem is to find self--consistently first--order changes $\delta \rho
({\bf r}t)$ and $\delta {\bf m}({\bf r}t)=\sum_{\nu }\delta m_{\nu }({\bf r}%
t){\bf e}^{\nu }$ induced by the field $\delta {\bf B}_{ext}({\bf r}t)$. If
the polarization $\delta {\bf b}$ in (\ref{f1}) is fixed to a particular $%
\mu $th direction, and $\delta {\bf m}({\bf r}t)$ is calculated afterwards,
a $\mu $th column of the spin susceptibility matrix $\chi _{\nu \mu }({\bf r}%
,{\bf q}+{\bf G},\omega )$ will be found \cite{Dia}. This essentially solves
the problem.

A central issue of employing {\em time--dependent }(TD)\ density functional
theory (DFT) \cite{TD-DFT} to find the quantities $\delta \rho ({\bf r}t)$
and $\delta {\bf m}({\bf r}t)$ is now discussed. The unperturbed density and
magnetization are described accurately by the static DFT and are expressed
via occupied Kohn--Sham states. This is by now a well established method in
practical {\it ab initio} calculations. In order to find the dynamical
response within TD DFT, only the knowledge of these unperturbed Kohn--Sham
states (both occupied and unoccupied) is required; no knowledge of real
excitation spectra (both energies and lifetimes) is necessary. This is the
main advantage of such approach. Unfortunately, within TD DFT, an accurate
approximation to the kernel $I_{xc}({\bf r},{\bf r}^{\prime },\omega )$
describing dynamical exchange--correlation effects is unknown while some
progress is currently been made \cite{ALDA}. In the following, the static
local density approximation (LDA) \cite{DFT} improved by a generalized
gradient approximation \cite{GGA} (GGA) is adopted to treat $I_{xc}({\bf r},%
{\bf r}^{\prime },\omega )$. To date, these are the most popular tools for
practical {\it ab initio} calculations, which are known to produce static
response functions as well as other ground--state, optical \cite{Rashkeev},
plus, recently \cite{EPI}, superconducting and transport properties for
large variety of solids in good agreement with experiments. The use of other
approximations to $I_{xc}({\bf r},{\bf r}^{\prime },\omega )$ will be
addressed in the future work.

An important issue of {\it variational} linear--response formulation is now
discussed. The advantage of variational principles for the calculation of
physical quantities is that if one makes a first--order error in the trial
function, the error in the variational quantity is of the second order. {\em %
Static} charge and spin susceptibilities appeared as second--order changes
in the total energy due to applied external fields can be calculated in a
variational way. This was demonstrated long time ago \cite{Vosko} on the
example of magnetic response, and, recently \cite{LR,Gonze}, in the problem
of lattice dynamics which is an example of charge response. The proof is
directly related to a powerful ''$2n+1$'' theorem of perturbation theory and
stationarity property for the total energy itself \cite{2n+1}. Any ($2n+1)$%
th change in the total energy $E_{tot}$ involves finding only ($n)$th order
changes in one--electron wave functions $\psi _{i}$, and corresponding
changes in the charge density as well as in the magnetization. Any ($2n$)th
change in $E_{tot}$ is then variational with respect to the ($n$)th--order
changes in $\psi _{i}$.

A time--dependent generalization of these results is now required. For TD
external fields, the action $S$ as a functional of $\rho ({\bf r}t)$ and $%
{\bf m}({\bf r}t)$ is considered within TD DFT \cite{TD-DFT,Liu}. These
functions are expressed via Kohn-Sham spinor orbitals $\vec{\psi}_{i}\left( 
{\bf r}t\right) $ satisfying TD Schr\"{o}dinger's equation \cite{Comm3}.
Therefore, $S$ as the stationary functional of $\vec{\psi}_{i}\left( {\bf r}%
t\right) $ is considered in practice$.$ When the external field is small,
the perturbed wave function is represented as $\vec{\psi}_{i}\left( {\bf r}%
\right) e^{-i\epsilon _{i}t}+\delta \vec{\psi}_{i}\left( {\bf r}t\right) $
and the first--order changes $\delta \vec{\psi}_{i}\left( {\bf r}t\right) $
define the induced charge density as well as the magnetization: 
\begin{eqnarray}
\delta \rho &=&\sum_{i}\left( \{\delta \vec{\psi}_{i}|I|\vec{\psi}_{i}\}+\{%
\vec{\psi}_{i}|I|\delta \vec{\psi}_{i}\}\right)  \label{f2} \\
\delta {\bf m} &=&\mu _{B}\sum_{i}\left( \{\delta \vec{\psi}_{i}|{\bf \sigma 
}|\vec{\psi}_{i}\}+\{\vec{\psi}_{i}|{\bf \sigma }|\delta \vec{\psi}%
_{i}\}\right)  \label{f3}
\end{eqnarray}
Here $\left\{ \left| {}\right| \right\} $ denotes averaging over spin
degrees of freedom only, $I$ is the unit 2$\times $2 matrix, and ${\bf %
\sigma }$ is the Pauli matrix. It is now seen that the knowledge of $\delta 
\vec{\psi}_{i}\left( {\bf r}t\right) $ will solve the problem.

In order to find $\delta \vec{\psi}_{i}\left( {\bf r}t\right) $, a
time--dependent analog of the ''$2n+1$'' theorem is now introduced. Any ($%
2n+1)$th change in the action functional $S$ involves finding only ($n)$th
order changes in the TD functions $\vec{\psi}_{i}\left( {\bf r}t\right) $,
and corresponding changes in charge density as well as in the magnetization.
Any ($2n$)th change in $S$ is then variational with respect to the ($n$%
)th--order changes in $\vec{\psi}_{i}\left( {\bf r}t\right) $. The proof is
the same as for the static case \cite{2n+1} if the stationarity property of $%
S$ and the standard TD perturbation theory are exploited. For important case 
$n=2$, this theorem makes the second--order change $S^{(2)}$ in the action 
{\em variational} with respect to the first--order changes $\delta \vec{\psi}%
_{i}\left( {\bf r}t\right) .$ If the perturbation has the form (\ref{f1}), $%
S^{(2)}$ is directly related to the real diagonal part of the
dynamical spin
susceptibility $Re[\chi _{\nu \mu }({\bf q}+{\bf G}^{\prime },{\bf q}+{\bf G}%
,\omega )]_{{\bf G}^{\prime }={\bf G}}$ , thus allowing its variational
estimate \cite{Liu}.

The problem is now reduced to find $S^{(2)}$ as a functional of $\delta \vec{%
\psi}_{i}\left( {\bf r}t\right) $ and to minimize it. This will bring an
equation for $\delta \vec{\psi}_{i}\left( {\bf r}t\right)$. Any change in
the action functional can be established by straightforward varying $S$ of
TD DFT \cite{TD-DFT,Liu} with respect to the perturbation (\ref{f1}). This
is analogous to what is done in the static DFT to derive, for example, the
dynamical matrix \cite{LR}. $S^{(2)}$ is found to be 
\begin{eqnarray}
&&S^{(2)}[\delta \vec{\psi}_{i}]=\sum_{i}2\left\langle \delta \vec{\psi}%
_{i}\left| H-i\partial _{t}I\right| \delta \vec{\psi}_{i}\right\rangle + 
\nonumber \\
&&\int \delta \rho \delta V_{eff}-\int \delta {\bf m}(\delta {\bf B}%
_{eff}+\delta {\bf B}_{ext})  \label{f4}
\end{eqnarray}
where the unperturbed 2$\times $2 Hamiltonian matrix $H=(-\nabla
^{2}+V_{eff})I-\mu _{B}{\bf \sigma B}_{eff}$. $V_{eff}$ and ${\bf B}_{eff}$
are the ground--state potential and magnetic field of the DFT. $\delta
V_{eff}$ and $\delta {\bf B}_{eff}$ are their first--order changes induced
by the perturbation (\ref{f1}) which involve the Hartree (for $\delta V_{eff}
$) and the exchange--correlation contributions expressed via $\delta \rho \,$%
and $\delta {\bf m}$ in the standard manner \cite{Callaway1975}.

The differential equation for $\delta \vec{\psi}_{i}\left( {\bf r}t\right) $
is now derived from the stationarity condition of\ (\ref{f4}). It is given
by 
\begin{equation}
(H-i\partial _{t}I)\delta \vec{\psi}_{i}+(\delta V_{eff}I-\mu _{B}{\bf %
\sigma }\delta {\bf B}_{eff})\vec{\psi}_{i}=0  \label{f5}
\end{equation}
This is a time--dependent version of the so--called Sternheimer equation
which is the Schr\"{o}dinger equation to linear order. It can be solved
easily on the frequency axis which substitutes $-i\partial _{t}$ by $%
\epsilon _{i}\pm \omega $ in (\ref{f5})$.$ The solution of the whole problem
assumes self--consistency: First, Eq. (\ref{f5}) is solved with the external
field (\ref{f1}). Second, $\delta \rho ({\bf r}\omega )$ and $\delta {\bf m}(%
{\bf r}\omega )$ are found according to (\ref{f2}) and (\ref{f3}). Third,
screened potential $\delta V_{eff}({\bf r}\omega )$ and magnetic field $%
\delta {\bf B}_{eff}({\bf r}\omega )$ are constructed. The cycle is repeated
again by solving (\ref{f5}). Evaluating $S^{(2)}$ after (\ref{f4}) brings
the variational estimate of the real diagonal susceptibility at the iteration.
The whole function is accessed via the knowledge of $\delta {\bf m}({\bf r}%
\omega ).$ The self--consistency should be done for every ${\bf q}+{\bf G}$
and $\omega $ value appeared in (\ref{f1}).

The advantages of this method are now seen: First, Eq. (\ref{f5}) does not
require an expansion of $\delta \vec{\psi}_{i}$ over complete set of
unperturbed wave functions $\vec{\psi}_{j}$ as it is done in the standard
perturbation theory. Only the knowledge of occupied and those unoccupied
states which are below $E_{F}+\omega $ is necessary. Second, the inversion
problem is substituted by the self--consistent finding of $\delta V_{eff}$
and $\delta {\bf B}_{eff}$. This normally requires about 10 iterations to
reach the convergency. Third, the method treats on the same footing both
longitudinal and transverse spin fluctuations which is achieved by choosing
the polarization $\delta {\bf b}$ of the external filed (\ref{f1}) along or
perpendicular to the magnetization axe. Fourth, the method gives an access
to {\em charge}--spin fluctuations via the knowledge of $\delta \rho ({\bf r}%
\omega ),$ and it is trivially converted to study dynamical charge
fluctuations, if a TD {\em scalar} filed of the type (\ref{f1}) is
considered as the perturbation.

An implementation of the method using linear muffin--tin orbital (LMTO)
representation is now discussed. As the original wave function $\vec{\psi}%
_{i}$ is expanded in terms of the LMTOs $\chi _{\alpha }$ with the
coefficients $\vec{A}_{i}^{\alpha }$, the first--order change $\delta \vec{%
\psi}_{i}$ generally involves both changes $\delta \vec{A}_{i}^{\alpha }$ in
the expansion coefficients and changes $\delta \chi _{\alpha }$ in the LMTO\
basis set \cite{LR}. Changes $\delta \vec{A}_{i}^{\alpha }$ are now new
variational parameters instead of $\delta \vec{\psi}_{i}.$ They must be
found by minimizing the functional (\ref{f4}). Changes $\delta \chi _{\alpha
}$ are, on the other hand, an auxiliary set of functions which is
constructed to make the expansion of $\delta \vec{\psi}_{i}$ fastly
convergent. Basis $\{\delta \chi _{\alpha }\}$ is normally adjusted to the
perturbation in the same way as the original basis $\{\chi _{\alpha }\}$ is
tailored to the unperturbed one--electron potential. Such perturbative
technique was found to be extremely efficient in the problem of lattice
dynamics \cite{LR}. In the magnetic response calculation introducing $\delta
\chi _{\alpha }$ is important for the fields exhibiting strong
short--wavelength oscillations. On the other hand, in the calculations with $%
{\bf G}=0$ in (\ref{f1}) the contributions originating from $\delta \chi
_{\alpha }$ are found to be small.

Numerical efficiency of the method is now demonstrated by calculating
spin susceptibilities for a number of metals. No shape approximations are
made in these calculations either for the charge densities and the
potentials or for the dynamical response functions. All the relevant
quantities are expanded in spherical harmonics inside muffin--tin spheres
and in plane waves in the interstitial region as it was done in original
full--potential and static linear--response LMTO methods \cite{LR}. The use
of GGA for exchange and correlation gives practically coinciding theoretical
and experimental lattice constants. Necessary Brillouin zone (BZ) integrals
are carried out using a multigrid tetrahedron technique \cite{LR} with
thousand ${\bf k}$ points.

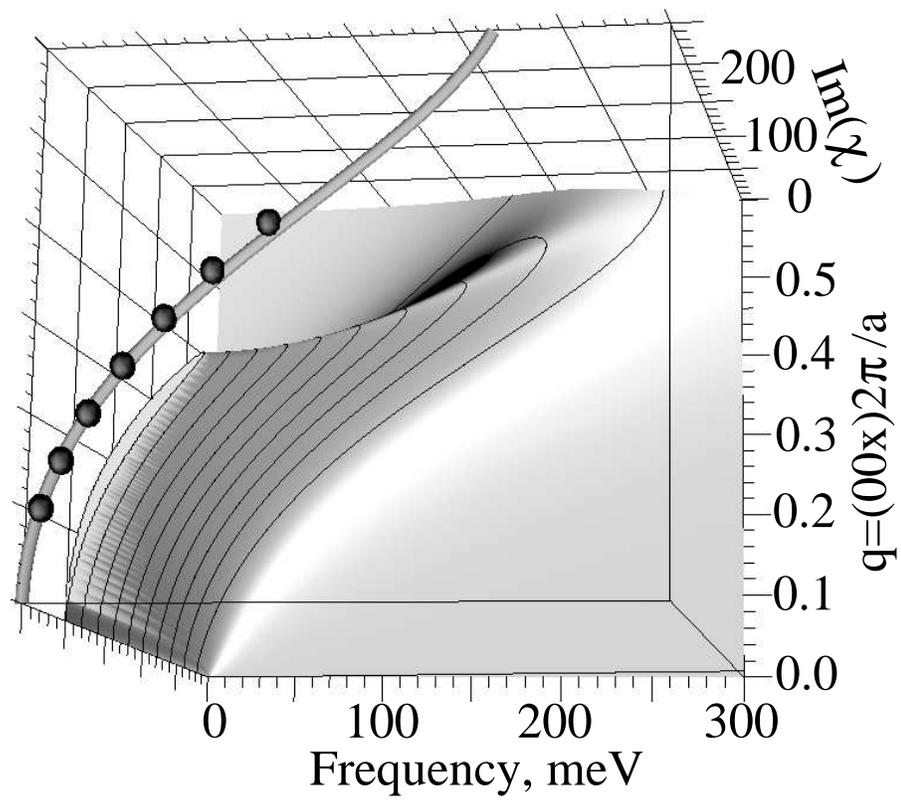
\begin{figure}[tbf]
\caption{
Calculated
$Im[\chi _{+-}({\bf q},{\bf q},\omega )]$ 
(arb. units) for Fe.
Top line shows the deduced magnon spectrum.
Balls indicate the experimental data [14].}
\label{Fe}
\end{figure}

The {\it ab initio }results obtained for bcc Fe are now reported.
Fig. \ref{Fe} shows calculated transverse spin susceptibility $Im[\chi _{+-}(%
{\bf q},{\bf q},\omega )]$ for ${\bf q}=(00x)2\pi /a$ \ At small ${\bf q}$
the undecaying spin waves are seen to persist in the structure of $Im[\chi ]$
exhibiting a standard dispersion law $\omega (q)=Dq^{2},$ where $D$ is the
spin stiffness of the material. The spin waves rapidly decay when ${\bf q}$
approaches to approximately one--half of the BZ. Similar picture has been
found for the ${\bf q}$'s along $(111)$ direction. The deduced spin--wave
spectrum is shown by the solid line on top of Fig. 1. It agrees well with
the experiment \cite{ExpFe} shown by circles as well as with the
recent frozen--magnon calculations \cite{Halilov}. Also, in agreement with
experiment any additional structure which can be attributed to the
appearance of optical spin--wave branches is not predicted. This advances 
the early RPA calculation \cite{CookeNiFe}.

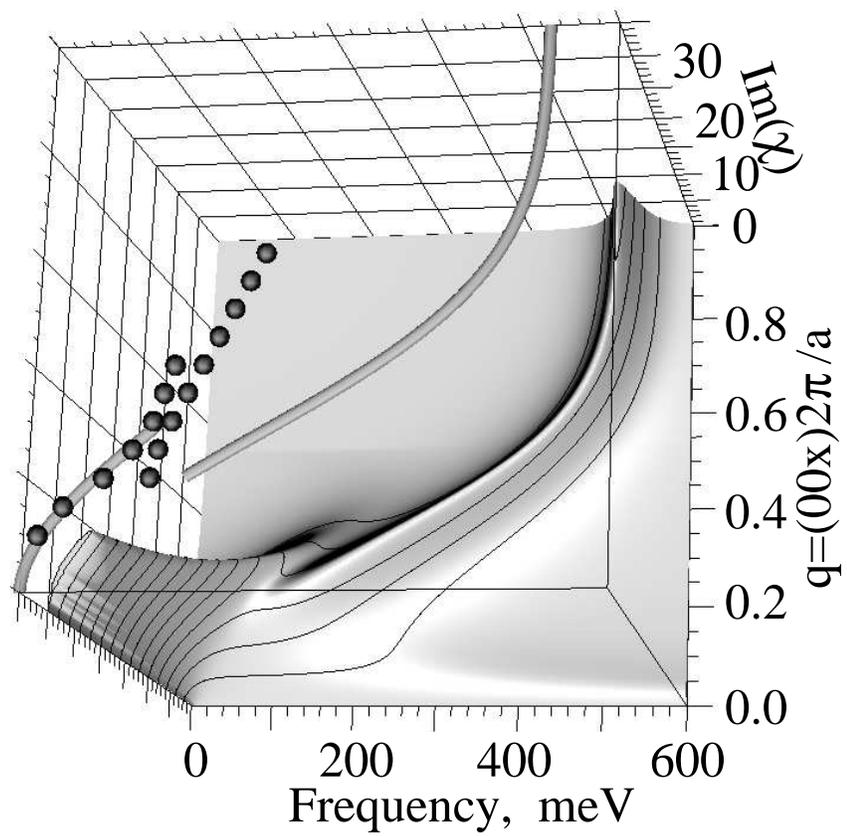
\begin{figure}[tbf]
\caption{
Calculated
$Im[\chi _{+-}({\bf q},{\bf q},\omega )]$ 
(arb. units) for Ni.
Top line shows the deduced magnon spectrum.
Balls indicate the experimental data [13].}
\label{Ni}
\end{figure}

$Im[\chi _{+-}({\bf q},{\bf q},\omega )]$ obtained
for fcc Ni is shown on Fig. \ref{Ni}. The unusual structure for the
energies near 100 meV and for the ${\bf q}^{\prime }s$ (0,0,0.2--0.4) 2$\pi /a
$ is clearly distinguishable. This was attributed to the appearance of the
optical branch in the spin--wave spectrum \cite{CookeNiFe,ExpNi}. However,
since this structure is seen to be only localized in a certain region of
${\bf q}$ space, its interpreting \cite{CookeNiFe}
as a well--defined branch persisting
to the BZ boundary might be not completely correct. The computations along
(111) direction do not show such unusual behavior. The obtained spin--wave
spectrum (line on top of Fig. 2) is in agreement with the measured
one (balls) \cite{ExpNi} in the low--frequency interval. 
However, a tendency to
overestimate spin--wave energies for higher $\omega $ is found both for
(001) and (111) directions. This is attributed to the poor treatment of
dynamical exchange--correlation effects due to simple GGA.

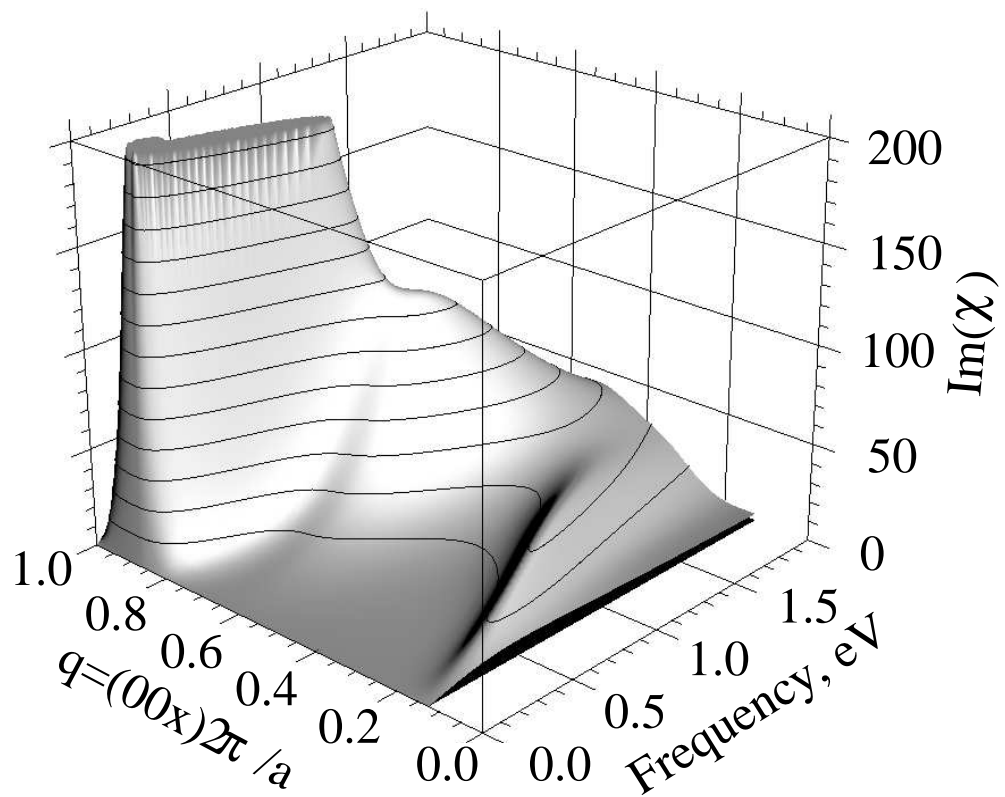
\begin{figure}[tbf]
\caption{Calculated
$Im[\chi ({\bf q},{\bf q},\omega )]$ (Ry$^{-1}$) for Cr.}
\label{Cr}
\end{figure}

Two examples of calculating paramagnetic spin fluctuations are now
considered. Fig. \ref{Cr} shows calculated $Im[\chi ({\bf q},{\bf q},\omega )]$ for
paramagnetic bcc Cr. A remarkable structure is clearly seen for the ${\bf q}%
^{\prime }$s near (0,0,$x_{SDW}\sim $0.9)2$\pi /a$, where the susceptibility 
is mostly enhanced at low frequencies (experimentally, 
$x_{SDW}$=0.95). This predicts Cr to be an incommensurate antiferromagnet. 
To clarify whether the Fermi--surface nesting is the origin of such behavior \cite{Cr},
the {\em non--interacting} susceptibility $Im[\chi _0({\bf q},{\bf q},\omega )]$
can be analyzed. 
It {\em does not} show up a structure peaked at $x_{SDW}\sim $0.9,
and is only a monotonically varying function when $x$ increases from 0 to 1.
This means that the generalized Stoner criterium $1=I_{xc}\chi _0({\bf q})$
does not necessarily assumes a peak in $\chi _0({\bf q}_{SDW})$ for Cr. 

$Im[\chi ({\bf q},{\bf q},\omega )]$ in Pd is found to be strongly enhanced
at small ${\bf q}^{\prime }s$ in complete agreement with the early studies 
\cite{WinterPdV}. Therefore, the method also confirms a closeness of Pd to
the ferromagnetic instability.

In conclusion, the developed approach is able to describe known
spin--fluctuational spectra of real materials which demonstrates its
efficiency for practical {\it ab initio} calculations. Also, more elaborate
approximations to the dynamical exchange and correlation are clearly
required in order to account for the observed discrepancies.

The author is indebted to O. K. Andersen, O. Gunnarsson, O. Jepsen, M. I.
Katsnelson, and A. I. Liechtenstein for many helpful discussions.

%

\begin{references}
\bibitem{Dia}  Diamagnetic response is not considered in this work.
\bibitem{Comm3}  To allow spin--orbit coupling, the Pauli equation should be
assumed here; $\vec{\psi}$ denotes a two-component spinor.
\end{references}
\end{document}